\documentstyle[12pt]{article}
\begin{document}

\noindent
{\Large\bf Some Quantum Aspects of Three-Dimensional 
Einstein-Chern-Simons-Proca
Massive \mbox{Gravity}}\\[3mm]

\noindent
{\sl C. PINHEIRO$^{(1)}$, G.O. PIRES$^{(2)}$ and N. TOMIMURA$^{(3)}$}

\ \\[2mm]

\noindent
\footnotesize{$^{(1)}$Departamento de F\'{\i}sica,
 Universidade Federal do Esp\'{\i}rito Santo --UFES, \\
Av. Fernando Ferrari, s/n 29060-900, Vit\'oria, ES, Brazil. \\
e-mail: fcpnunes@cce.ufes.br \\

\noindent
$^{(2)}$Instituto de F\'{\i}sica, Universidade Federal do Rio  de
Janeiro --
UFRJ, \\ Caixa Postal 68528, Cep 21945-970, Rio de Janeiro, RJ, Brazil. \\

\noindent
$^{(3)}$Instituto de F\'{\i}sica, Universidade Federal Fluminense -- UFF,
Outeiro de S\~ao Jo\~ao Batista, s/n$^{\underline{0}}$, Niter\'oi, RJ, Brazil.}

\ \\[0.5cm]
\begin{center}
\begin{minipage}{12cm}
{\bf Summary.} -- We present a 3-dimensional model for 
massive gravity with masses induced
by topological (Chern-Simons) and Proca-like mass terms. 
Causality and unitarity
are discussed at tree-level. Power-counting renormalizability 
is also contemplated.
\end{minipage}
\end{center}

\normalsize
\vspace{2cm}
\paragraph*{}
It is well-known that the linearized Einstein Lagrangian 
is the only one that describes
(consistently with locality, causality and tree-level unitarity) a massless
rank-2 symmetric tensor field, $h_{\mu\nu}$, in four space-time 
dimensions \cite{1}.
However, a question arises that regards the massive version of this symmetric
field: is the Pauli-Fierz Lagrangian the best way to describe 
the propagation
of such massive fields?

In the work of ref. [1], van Nieuwenhuizen presents a very detailed discussion
on linearized gravitation and Lagrangian densities for rank-2 tensor fields.
There, he treats very throughly the issues of gauge symmetry, ghosts and the introduction
of mass for this type of fields. 

Gravitational theories are generally formulated for free 
fields that exhibit
a definite non-negative energy (absence of ghosts) and, 
in the massive case, one
expects that the local character of the particular model 
be consistent, in that no tachyon
shows up. Therefore, the poles in the propagators of a massive theory must 
necessarily be associated to real masses; imaginary or higher-order poles 
invalidate the physical consistency of the theory under study.

A very straightforward way of deriving propagators is by 
means of the inversion
of the field equation. Here, to accomplish such a procedure, 
we shall work 
with the extended version of the Barnes-Rivers spin 
operators \cite{2}. By extended,
we mean that a set of new operators \cite{3,4} has been 
adjoined to the operators
originally studied by Barnes and Rivers \cite{2}, in order 
to establish a formalism that 
encompasses the case of 3-dimensional topologically massive 
gravity \cite{4}.

Once a unitary and causal gravitational model has 
been formulated (at least 
at the level of the tree amplitudes), a further 
problem one has to face is the 
study of its renormalizability. The presence of 
infinities in the realm of quantum
field theories is a current and problematic question: 
gravitation is not an exception
to this matter, at least in 4 dimensions, as far as one 
sticks to unitarity.

The main purpose of this work is to present a 
3-dimensional model for 
(Proca-like) massive gravity  
in the presence of a Chern-Simons mass term, 
with the requirements 
of tree-level
causality, unitarity and power-counting 
renormalizability.

With the reality and conservation of the 
energy-momentum tensor 
(as discussed in refs. 
[1,3,5]), the tree-level unitarity will be ensured. 
Also, the explicit calculation
of the residues of the propagator at the simple poles 
shall guarantee that negative-norm
states (ghosts) decouple form physical amplitudes.

We begin by adopting the following Lagrangian 
for a theory of massive gravity 
in 3 dimensions, with the Chern-Simons 
topological term:

\begin{equation}
L = L_{H.E} + L_{P} + L_{CS}\ .
\end{equation}

The first term is the usual Einstein-Hilbert Lagrangian, given by

\begin{equation}
L_{H.E} = -\frac{1}{2\kappa^2} \sqrt{-g} R,
\end{equation}
where $\kappa$ plays the role of the coupling constant for gravity.

The second one is the Proca mass term, namely

\begin{equation}
L_p = -\frac{1}{4}\ m^2 (h^{\mu\nu}h_{\mu\nu} - \xi h^{\mu}_{\mu}
h^{\nu}_{\mu})\ ,
\end{equation}
where $m$ and $\xi$ are parameters in terms of which the physical 
masses will be calculated.

Finally, the last term, known as the to\-po\-log\-i\-cal 
Chern-Simons La\-grang\-i\-an, 
exhibits the following form:

\begin{equation}
L_{C.S} = \frac{1}{\mu} \varepsilon^{\lambda \mu \nu} \Gamma^{\rho}_{\lambda\sigma} 
\left(\partial_{\mu}\Gamma^{\sigma}_{\rho\nu} +
\frac{2}{3} \Gamma^{\sigma}_{\mu\varphi}
\Gamma^{\varphi}_{\nu\rho}\right)\ ,
\end{equation}
where $\mu$ is a dimensionless parameter that, along with $m$ and $\xi$, determines the
physical masses, as read off from the $h_{\mu\nu}$ - field propagator.

The metric used in (1) is the linearized one:

\begin{equation}
g_{\mu\nu}(x) = \eta_{\mu\nu} + \kappa h_{\mu\nu} (x) .
\end{equation}

Taking into consideration the formalism of spin 
operators by Barnes and Rivers \cite{2} and 
its extended algebra \cite{3}, then the free 
part of the Lagrangian (1) can be
cast into a bilinear form as below:

\begin{equation}
L = \frac{1}{2} h^{\mu\nu} O_{\mu\nu, \kappa\lambda}h^{\kappa\lambda}\ ,
\end{equation}
where $O_{\mu\nu , \kappa\lambda}$ is an operator 
written in terms of the spin-projector operators in the
space of rank-2 symmetric tensors, according to the expression

\begin{eqnarray}
&&O_{\mu\nu ,\kappa\lambda} = -\frac{1}{2} (\Box + m^2) P^{(2)} -
\frac{m^2}{2} P^{(1)}_m +
\left[\Box - 
\frac{m^2}{2} (1-3\xi )\right]P_s^{(0)} + \nonumber \\
&& -\frac{m^2}{2} (1-\xi ) P_w^{(0)} + 
\frac{\sqrt{3}}{2} m^2\xi (P_{sw}^{(0)} + P_{ws}^{(0)}) +
\frac{1}{2M} (S_1+S_2) , 
\end{eqnarray}
where
\begin{eqnarray}
&&P^{(2)}_{\mu\nu ,\kappa\lambda} \equiv \frac{1}{2} (\theta_{\mu\nu}\theta_{\kappa\lambda} 
+ \theta_{\mu\lambda}\theta_{\nu\kappa}) - \frac{1}{3} \theta_{\mu\nu}\theta_{\kappa\lambda} , \nonumber \\
&& P^{(1)}_{\mu\nu ,\kappa\kappa}\equiv \frac{1}{2} (\theta_{\mu\kappa}\omega_{\nu\lambda} + 
\theta_{\mu\lambda}\omega_{\nu\kappa} + 
\theta_{\nu\kappa}\omega_{\mu\lambda} + \theta_{\nu\lambda} 
\omega_{\mu\kappa}) , \\
&& P^{(0)}_{s\mu\nu ,\kappa\lambda} \equiv \frac{1}{3}
\theta_{\mu\nu}\theta_{\kappa\lambda}   ,
\nonumber \\
&& P^{(0)}_{w \, \mu\nu ,\kappa\lambda} \equiv \omega_{\mu\nu}\omega_{\kappa\lambda} , \qquad
P^{(0)}_{sw \, \mu\nu ,\kappa\lambda} \equiv \frac{1}{\sqrt{3}} \theta_{\mu\nu}\omega_{\kappa\lambda},
\nonumber \\
&& P^{(0)}_{ws \, \mu\nu ,\kappa\lambda} \equiv \frac{1}{\sqrt{3}}
\omega_{\mu\nu}\theta_{\kappa\lambda},\nonumber \\
&& S_{1 \, \mu\nu ,\kappa\lambda} \equiv
\frac{(-\Box )}{4} \left\{\varepsilon_{\mu\alpha\lambda} \partial_{\kappa}
\omega^{\alpha}_{\nu} + \varepsilon_{\mu\alpha\kappa} 
\partial_{\lambda}\omega^{\alpha}_{\nu} +
\varepsilon_{\nu\alpha\lambda}\partial_{\kappa}\omega^{\alpha}_{\mu} +
\varepsilon_{\nu\alpha\kappa} \partial_{\lambda} \omega^{\alpha}_{\mu}\right\}, \nonumber
\\
&&S_{2 \, \mu\nu ,\kappa\lambda} \equiv \frac{\Box}{4} \left\{\varepsilon_{\mu\alpha\lambda} 
\eta_{\kappa\nu} + \varepsilon_{\mu\alpha\kappa} \eta_{\lambda\nu} 
+\varepsilon_{\nu\alpha\lambda}\eta_{\kappa\mu} +
\varepsilon_{\nu\alpha\kappa}\eta_{\lambda\mu}\right\}\partial^{\alpha} ,
\nonumber
\end{eqnarray}
$\theta_{\mu\nu}$ and $\omega_{\mu\nu}$ being the usual transverse and longitudinal
projectors on the space of vectors.

It is worthwhile to notice that  these operators are not true 
projectors in that 
they are not all idempotent. They are simply operators 
that form a basis in spin
space.

In the expression above, $M$ is a simple redefinition of the
Chern-Simons mass parameter,
$M=\frac{\mu}{8\kappa^2}$. It is easy to observe that for $\xi =1$ and $M\rightarrow \infty$, one immediately reobtains the Einstein-Proca
theory in D dimensions \cite{3}.

The propagator of the theory is given by the expression

\begin{equation}
\langle T[h_{\mu\nu}(x)h_{\kappa\lambda} (y)]\rangle = iO^{-1}_{\mu\nu , \kappa\lambda} 
\delta^3 (x-y) 
\end{equation}
with $O^{-1}$ being the inverse operator of (7). 
To calculate such an operator, one has
to use the technique proposed by Barnes and Rivers \cite{2} 
with the difference
now that the new operators that appear (namely $S_1$ and $S_2$) are due to the topological
Lagrangian of eq. (4). The details involved in the task of 
deriving the explicit
form of $O^{-1}$ are described in  the work of ref. [3].

The operator $O^{-1}$ can therefore be explicitly written as

\begin{equation}
O^{-1}_{\mu\nu ,\kappa\lambda} = \sum^8_{i=1} X_i P^{(i)}_{\mu\nu , \kappa\lambda} \, , 
\end{equation}
where $P^{(i)}$ denotes the whole set of spin operators, whereas the $X_i$'s
stand for unknown coefficients. The latter can be found to read as below:

\begin{eqnarray}
&&X_1 = -\frac{2M^2(\Box +m^2)}{\Box^2M^2+2\Box m^2M^2+m^4M^2+\Box^3}, \nonumber \\
&&X_2 = -\frac{2}{m^2}, \\
&& X_3 = \frac{E}{F},\nonumber
\end{eqnarray}
where\\ $E=[(-3+3\xi )\Box^3 + (-4M^2+4\xi M^2)\Box^2 + 
(4m^2\xi M^2 - 6m^2M^2)
\Box - 2m^4M^2]$\\ and \\ $F=[(-1+\xi )\Box^4 + (m^2-3m^2\xi -M^2+\xi M^2)
\Box^3 + (-m^2M^2-m^2\xi M^2) \Box^2 + 
+(m^4M^2-5m^4M^2\xi )\Box -3\xi m^6M^2 + m^6M^2]$,

\begin{eqnarray}
&& X_4 = 2\left[\frac{2m^2\xi + \Box -m^2}{m^2[(\xi -1)\Box -3m^2\xi +m^2]}\right],
\nonumber \\
&& X_5 = -\frac{2\sqrt{3} \xi}{\Box \xi -3m^2\xi -\Box + m^2}, \nonumber \\
&& X_6 =  -\frac{2\sqrt{3} \xi}{\Box \xi -3m^2\xi -\Box + m^2}, \nonumber \\
&& X_7 =  -\frac{2M}{\Box^2M^2 +2\Box m^2M^2+m^4M^2+\Box^3}, \nonumber \\
&& X_8 =  -\frac{2M}{\Box^2M^2 +2\Box m^2M^2+m^4M^2+\Box^3}. \nonumber 
\end{eqnarray}

If we consider $\xi =1$ and $M\rightarrow \infty$ in eq. (9), we reobtain
the propagator for massive Einsten-Proca gravity in 3 space-time dimensions.

In the following, we are going to discuss the unitarity of the Einstein-Chern-Simons-Proca
model. As done in \cite{3,5}, here too, we proceed to this analysis 
by studying
the properties of the current-current transition amplitude,

\begin{equation}
A = T^{\mu\nu} (-k)< h_{\mu\nu}(-k)h_{\kappa\lambda}(k) > T^{\kappa\lambda}(k) ,
\end{equation}
where $T_{\mu\nu}$ is the energy-momentum tensor and the expression between parentheses
is the propagator of the eq. (9) written in momentum space.

Taking into consideration the reality condition

\begin{equation}
T_{\mu\nu}(-k)=(-1)^{\delta_{\mu 4}+\delta_{\nu 4}} (T_{\mu\nu}(k))^{*} ,
\end{equation}
and the orthogonality condition between the energy-momentum tensor and the longitudinal operator
$\omega_{\mu\nu}$ present in all the spin operator projectors $P^{(i)}$, 
except $P^{(2)}$ and $P^{(0)}_s$, we can estimate the definite 
positivity of the
residue at the poles.

In a concise way, the transversality condition is given by

\begin{equation}
\omega T = 0 .
\end{equation}
From now on, our aim is to analyse the sign of the imaginary part of the residues
of the amplitude (12) at the poles of the $h_{\mu\nu}$ - field propagator; 
to ensure a healthy particle spectrum, we must find out that

\begin{equation}
Im\ Res \ A >0 ,
\end{equation}
for any pole in $k^2$.

To do so, we need the poles of the coefficients $X_1$ and $X_3$, respectively.
This is so because they are the only ones to survive the transversality of $T^{\mu\nu}$ in eq. (12).

We pick out the particular case where $\xi =1$, leaving the parameters
$m$ and $\mu$ free. In so doing, one can readily check that the coefficients of $P^{(2)}$ 
and $P^{(0)}_s$ display the same structure of poles, 
namely, the roots of the cubic equation

\begin{equation}
(k^2)^3 - M^2(k^2)^2 + 2m^2 M^2k^2 - m^4M^2 = 0 .
\end{equation}
The solutions to (16) are more easily discussed in 
terms of the ratio of masses,

\begin{equation}
\alpha \equiv  \frac{M}{m}.
\end{equation}

It can be checked that the condition

\begin{equation}
\alpha > \frac{3\sqrt{3}}{2}
\end{equation}
must be fulfilled in order that poles that correspond to 
tachyons and ghosts be
suppressed from the spectrum. The absence of tachyons 
is ensured as long as the poles
in $k^2$ are all positive; as for the ghosts, they do 
not show up since it turns
out that

\vspace{0.5cm}
\[ \hspace{6cm} Im\ Res \ A >0 , \hspace{4cm} \]

\vspace{0.5cm}
\noindent
for each of the simple poles of the 
$h_{\mu\nu}$-field propagator. 
Also, though it is not 
immediately clear, the condition (18) automatically 
avoids the apperance of higher-order
poles, which would plague the spectrum with ghosts.

For example, if we choose to take $\alpha =3$, 
the folllowing roots are obtained:

\begin{eqnarray}
&& k_1^2 = 6.4114 m^2 ,\nonumber \\
&& k^2_2 = 0.7737 m^2 , \\
&& k^2_3 = 1.8154 m^2 . \nonumber    
\end{eqnarray}

Going back to expression (15) with the poles given in (19), 
one can verify that the
three massive gravitons (19) are in fact the {\it mediators} 
of gravitation in (2+1), for,
at each of these poles, the imaginary part of the residue 
of the current-current
transition amplitude is positive-definite, which guarantees 
the unitarity of the
model at tree-level in 3 dimensions.

The coefficients $X_1$ and $X_3$ of the propagator signal 
the renormalizability of the 
theory, as there only appear integrals of the 
form $\int d^3k \frac{1}{k^4}\sim
\frac{1}{k}$ throughout loop calculations; this indicates 
a non-divergent theory
for asymptotic values of momenta. This simply supports 
the results on the renormalizability
of Chern-Simons gravity, according to the results of ref. [6].

We conclude by stating that Einstein-Chern-Simons-Proca
theory is a reasonable
theory in 3 space-time dimensions: it is causal, 
unitary and renormalizable.

\vspace{.5cm}

The authors wish to thank Dr. J.A. Helay\"{e}l-Neto for 
discussions and critical
reading of the manuscript. Thanks are also due to Mr. C. Sasaki for 
the invaluable
support with computer algebra. The authors are partially supported by the 
Conselho Nacional de Desenvolvimento Cient\'{\i}fico e  Tecnol\'ogico, 
CNPq-Brazil.



\begin{thebibliography}{99}
\bibitem{1} P. Van Nieuwenhuizen, Nucl. Phys. B{\bf 60} (1973) 478.
\bibitem{2} R.J. Rivers, Il Nuovo Cimento {\bf 34} (1964) 387; K.J. Barnes, Ph.D. Thesis (1963), unpublished.
\bibitem{3} C. Pinheiro and  G.O. Pires, Phys. Lett. B {\bf 301} (1993) 339; C.
Pinheiro, G.O. Pires and F.A.B. Rabelo de Carvalho, EuroPhys. Lett., {\bf 25}
(1994) 329.
\bibitem{4} Deser, R. Jackiw and S. Templeton,  Ann. Phys. (N.Y.) {\bf 140} 
(1982) 372.
\bibitem{5} Carlos Pinheiro, Gentil O. Pires and Claudio Sasaki. ``On a Three-Dimensional Gravity
Model with Higher Derivatives'', to be published in Gen. Relativ.  Grav.; 
Carlos Pinheiro, Gentil O. Pires and 
Fernando A.B. Rabelo de Carvalho, ``Some Quantum Aspects
of D=3 Space-Time Massive Gravity'', to be published in Braz. J. Phys.
\bibitem{6} Deser and Z. Yang, Class. Quantum Grav. {\bf 7} (1990) 1603.
\end{thebibliography}
\end{document}